\documentclass{JAC2003}

\usepackage{graphicx}
\usepackage{epsfig}

\newcommand{\AmS}{{\protect\the\textfont2
  A\kern-.1667em\lower.5ex\hbox{M}\kern-.125emS}}

\begin{document}
\title{Physics at a \boldmath {$\gamma\gamma $, $e\gamma$} and 
{\boldmath $e^-e^-$} Option for a 
Linear Collider}

\author{A. De Roeck\thanks{
The work reported in this talk was done by the members
of the ``$\gamma\gamma $ and $e\gamma$  physics''
working group of the Extended ECFA/DESY Study;
D. Anipko (Nowosibirsk),
E. Asakawa  (Tokyo U.),
D. Asner (Cornel),
I. Bozovic (VINCA Belgrade),
W. Da Silva (Paris VI),
A. De Roeck (CERN),
A. Finch (Lancaster),
I. Ginzburg (Nowosibirsk),
R. Godbole (Bangalore),
J. Gronberg (LLNL Livermore),
C. Heusch (Santa Cruz),
G. Klemz (DESY-Zeuthen),
M. Kr\"amer (Edinburgh),
F. Kraus (Dresden),
M. Krawczyk (Warsaw),
J. Kwiecinski (Krakow),
V. Makarenko (NC PHEP Minsk),
I. Marfin (NC PHEP Minsk),
S. Maxfield (Liverpool),
D. Miller (CERN),
K. Moenig (DESY-Zeuthen),
M. M\"uhlleitner (PSI),
F. Nagel (Uni Heidelberg),
P. Niezurawski (Warsaw),
A.V. Pak (Novisibirsk),
D.V. Pavluchenko (Novisibirsk),
S.S. Petrosyan (Novisibirsk),
A. Rosca DESY-Zeuthen),
S. Schumann (Dresden),
J. Sekaric (DESY Zeuthen),
V.G. Serbo (Novisibirsk),
T. Shishkina (NC PHEP Minsk),
S. Soeldner-Rembold (Manchester),
A. Stahl (DESY Zeuthen),
V. Telnov (Novisibirsk),
M. Velasco (Northwestern),
M. Wing (Bristol Univ. \& DESY),
A.F.\.Zarnecki (Warsaw),
},
CERN, 1211 Geneva 23, Switzerland}

\def\GG{\gamma\gamma}
\def\GE{e\gamma}
\def\eG{e\gamma}
\def\EPEM{$e^+e^-$}
\def\CPbar{\hbox{{\rm CP}\hskip-1.80em{/}}}

\newcommand{\ggx}{{$\gamma\gamma$}}
\def\sigg{{$\sigma_{\gamma\gamma}^{tot}$}}
\def\fg{{$F^{\gamma}_2$}}
\def\xg{{$x_{\gamma}$}}
\newcommand{\syy}{{$\sqrt{s}_{\gamma\gamma}$}}
\def\gstar{{$\gamma^*\gamma^*$}}
\def\ep{{e$^+$e$^-$}}

\maketitle

\def\syy{{$\sqrt{s}_{\gamma\gamma}$}}

\begin{abstract}
This report presents a review of the studies made in the 
 working group on ``$\gamma\gamma$ and $ e\gamma$ physics" 
of the ECFA/DESY workshop on 
linear collider
physics.
It reports on several new physics studies, in particular s-channel Higgs 
production. A summary of R\&D activities for the interaction 
region is presented. 
The merits of $e^-e^-$ collisions are briefly recalled.
\vspace{1pc}
\end{abstract}


\section{Introduction}
A future  $e^+e^-$ linear collider (LC) offers 
excellent new opportunities for 
the study of high energy particle collisions. 
The idea  to convert the 
electron beams of a LC into photon beams, by laser backscattering,
and thus create a photon collider (PC), was first discussed about 
20 years ago in \cite{first}.
Projects for a future LC collider are studied in Europe (TESLA,CLIC),
the US (NLC) and Asia (JLC), and all consider a PC as a possible
additional option.
Recently, in the context of the 
ECFA-DESY LC study, a detailed discussion of the physics and design of a PC 
was presented in the TESLA-TDR~\cite{pctdr}
and in~\cite{golden}. This paper reviews the work done during the 
last two years in the study 
group ``$\gamma\gamma $ and $e\gamma$ physics''  
 of the extended  ECFA/DESY workshop on physics and detectors at a 
linear collider.

A plethora of new
and exciting measurements 
become accessible with a PC, in particular 
Higgs boson studies, but also  searches for new physics and electroweak,
top  and QCD
measurements
can be made often in a complementary way compared to $e^+e^-$ 
collisions. The precision reached at a PC is competitive if 
sufficiently high luminosities can be reached.

Examples of advantages of a PC  include:
\begin{itemize}
\item Higher cross sections for charged particles than in $e^+e^-$.
\item Different $J^{PC}$ states than for $e^+e^-$.  
\item Higgs can be s-channel produced as a resonance.
\item CP analysis opportunities for Higgs bosons
\item Precise test of  the coupling to photons
\item Possible higher mass discovery range for e.g.~$H, A$, and sleptons
\end{itemize}
Note that a PC needs no positron drive beam but electron beams, 
which can be produced with relatively high polarisation, are sufficient.

\begin{figure}[tbh]
\begin{center}
{\includegraphics[width=0.8\linewidth]{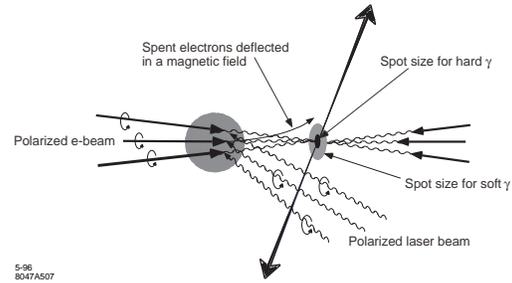}}
\caption{\label{fig:contr_02} 
 A sketch of the creation of a photon beam by Compton backscattering 
 of laser photons off  beam electrons.
 }
\end{center}
\end{figure}

The proposed technique for a PC consists of using laser backscattering
as shown in Fig.~1.
A
low energy (typically 1 eV) laser beam of photons collides with the 
high energy (typically 250-500 GeV) electron beam and is backscattered 
receiving a major fraction of the incoming electron energy.
The maximum energy of the generated photons is 
given by $E_\gamma^{max} = xE_e/(1+x)$, with $E_e$ the electron beam energy
and $x = 4E_eE_L\cos^2(\theta/2)/m_e^2c^4$ with $E_L$ and $\theta$ the laser 
photon energy and angle between the electron and laser beam.
The distance of the conversion to the interaction point is in the range 
of several mm. A typical value for $x$ is 4.8, which leads 
to photon spectra which peak around $0.8E_e$. 
 The energy distribution depends on the 
polarisation of the photon ($P_c$) and electron beam ($\lambda_e$), 
 the most peaked spectrum is obtained when
$P_c \lambda_e =-1$. 
In reality, due to the maximum polarizability of the electron beam 
a value close to $P_c \lambda_e =-0.8$ can be reached.
Sometimes it is advantageous to have a broader spectrum, e.g. to discover 
particles with unknown masses, in which case the configuration 
$P_c \lambda_e =+0.8$ will be more useful.

The polarization of both beams can 
be further used to produce interactions with the same ($J_z=0$) or 
opposite ($J_z=2$)  photon helicities, useful e.g.~for Higgs studies.
 Higher geometrical luminosities
 can be achieved for photon colliders than 
 for genuine $e^+e^-$ colliders, due to the absence or strong reduction of
 beamstrahlung in the interaction region.
  The 'luminosity'
 is usually defined to be 
 the luminosity corresponding to the  region 
$ \sqrt{s_{\gamma\gamma}} > 0.8 \sqrt{s_{\gamma\gamma, max}}$
and is typically 
 10\% of the geometrical $e^+e^-$ luminosity. For the TESLA  parameters, but
 including
a smaller horizontal $\beta$ function at the interaction point
namely 1.5 mm in $x$, compared to 15 mm for the $e^+e^-$ beam design, 
and reducing the horizontal emittance from 553 nm  to 140 nm, leads to 
 $L_{\gamma\gamma}(\sqrt{s_{\gamma\gamma}}>0.8 \sqrt{s_{\gamma\gamma, max}})
\sim \frac{1}{3}L_{e+e-}$.
This gives event samples corresponding to
$O(100)$ fb$^{-1}$ per year for the PC. 
A PC
\begin{itemize}
\item needs a second interaction point
\item needs a cross angle
\item has a rather peaked but somewhat smeared centre of mass system
(CMS) energy spectrum
\end{itemize}

Both high energy $e\gamma$ and $\gamma\gamma$ interactions can be 
provided, depending on whether only one or both lepton beams are converted.

%
%



 \par

\section{tools}
During this workshop major progress was made on the  
development and completion of the tools to study physics at a 
$\gamma\gamma$ collider. These tools have now reached 
 a high level of  maturity.

\begin{figure}[tbh]
\begin{center}
{\includegraphics[bbllx=0pt,bblly=0pt,bburx=570pt,bbury=310pt,width=0.9\linewidth]{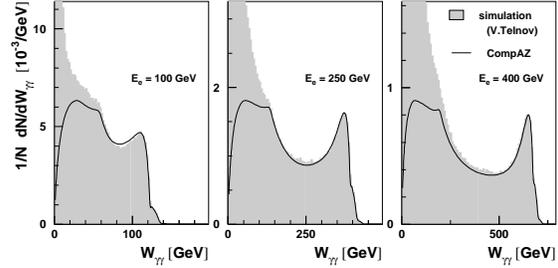}}
\caption{\label{fig:compaz} 
Comparison of the center of mass  energy distribution obtained from full
simulation of the luminosity spectrum~\cite{phocol} with results from CompAZ,
for three electron beam energies~\cite{compaz}}
\end{center}
\end{figure}

Luminosity spectra at photon colliders can not be described completely by 
 effective photon spectra due to the energy-angle correlation in Compton 
scattering
 and beam
collision effects. Fully detailed luminosity distributions were obtained by  
a complete simulation
 of beam collisions, resulting in 'collision events' that contain
the 
types of colliding particles (photon, electron, positron), 
their energies and polarizations.
The  PHOCOL 
program~\cite{phocol} was used to generate these 
collision events for several $e^-e^-$ CMS energies and laser 
configurations. 
PHOCOL includes non-linear corrections and contributions
of higher order processes. An example of a 
$\gamma\gamma$ CMS energy distribution 
is  shown in Fig.~\ref{fig:compaz}.
The event files can be used by the CIRCE program~\cite{circe}.
These luminosity spectra are also used to tune a simple model based 
on analytical
formulae for the Compton scattering (CompAZ~\cite{compaz}). The results 
of 
such a tune
are shown in Fig.~2 as well.
While being an approximation, these spectra are nevertheless extremely 
convenient for studies e.g. at different energies other than
the (few) ones for which event files were produced.

A version of the fast detector simulation package
SIMDET, including modifications for the PC 
interaction point (IP) has been used.
Overlap events from the QCD background can be added to the signal events. For  
TESLA luminosities,  we expect typically on average about one overlaying 
event 
at low energy ($\sqrt{s_{ee}} \sim  200$ GeV, also called the Higgs mode
since it would be best suited  for the study of a light Higgs with
mass $\sim 120$ GeV) and two  events at 
nominal energy ($\sqrt{s_{ee}} \sim 500 $ GeV).

Background studies~\cite{moenig} have been made for incoherent and 
coherent $e^+e^-$ pair production.
A new two-mask design in the IP reduces the background by a factor 2-3 with
respect to 
the previous layout; the details are still being optimized. 
Tracks in the TPC and hits in the vertex detector from 
incoherent and coherent pairs were
found to be tolerable and similar to the expected background at 
an $e^+e^-$ collider interaction region. Hence there is now 
evidence that a similar vertex detector as for an $e^+e^-$ collider
detector can be used 
for  a PC detector, and therefore a similar quality in $b$-tagging 
can be achieved.
The neutron background  is still under study but the first results 
show that it is tolerable
as well~\cite{neutron_telnov}.

During this workshop we also had direct contact with 
MC developers which resulted in getting requirements implemented in e.g.~the 
new SHERPA generator~\cite{kraus}, and getting good MC parameter
tunes
for PYTHIA and HERWIG (using mostly HERA $\gamma p$ data) from the JETWEB 
team~\cite{wing}.

On the web page of the working group a link directing to the page with the 
tools can
be found:
http://www-h1.desy.de/$\,\tilde{}\,$maxfield/ggcol/lcgg.html.

\begin{figure}[tbh]
\begin{center}
{\includegraphics[bbllx=0pt,bblly=70pt,bburx=420pt,bbury=420pt,width=0.9\linewidth]{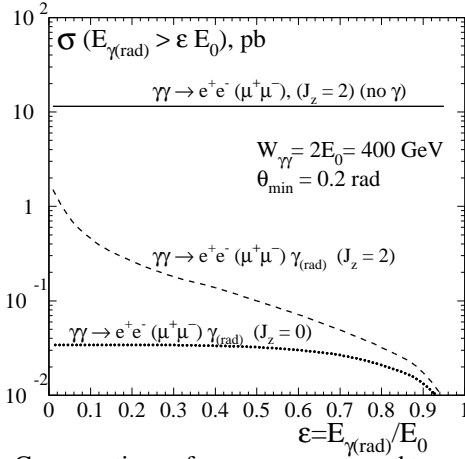}}
\caption{\label{fig:lumi} 
Cross sections of processes proposed to measure luminosity at a $\gamma\gamma$ 
collider~\cite{telnov_lum}.}
\end{center}
\end{figure}

\section{Luminosity}
One of the topics studied in detail is the precision with which
the luminosity  can be
measured. The following processes are 
proposed for the $\gamma\gamma$ mode~\cite{moenig_lum,telnov_lum}:
\begin{itemize}
\item $ee \rightarrow ee \, (\mu\mu)$
\item $ee \rightarrow ee\gamma \, (\mu\mu\gamma)$
\item $ee \rightarrow $ 4 leptons
\end{itemize}

The cross sections for these channels are shown in Fig.~\ref{fig:lumi}.
The first channel can give the highest precision $\sim 0.1\%$ (stat)
but cannot be used for $J_z = 0$, i.e. for the Higgs study, because it is 
suppressed as $m^2_l/s$, with $m_l$ the lepton mass.
In that case, however,  the second channel can be used. For two years of running
the statistical precision for the channel $ee \rightarrow ee\gamma$, 
using realistic detector cuts, is 
\begin{equation}
\frac{\Delta L}{L}(\sqrt{s} 
>0.8 \sqrt{s_{\gamma\gamma, max}}) = 0.4 \% 
\end{equation}

\begin{equation}
\frac{\Delta L}{L}(m_H\pm 2 {\rm GeV}) = 1.0 \% 
\end{equation}

For $e\gamma$ collisions the following processes are suggested:
\begin{itemize}
\item $e\gamma \rightarrow e\gamma, eZ$
\item $e\gamma \rightarrow eee$.
\end{itemize}

The statistical precision that can be achieved 
is better than 1\% for one 
year of running.

\begin{figure}[tbh]
\begin{center}
{\includegraphics[bbllx=0pt,bblly=120pt,bburx=550pt,bbury=680pt,width=0.85\linewidth]{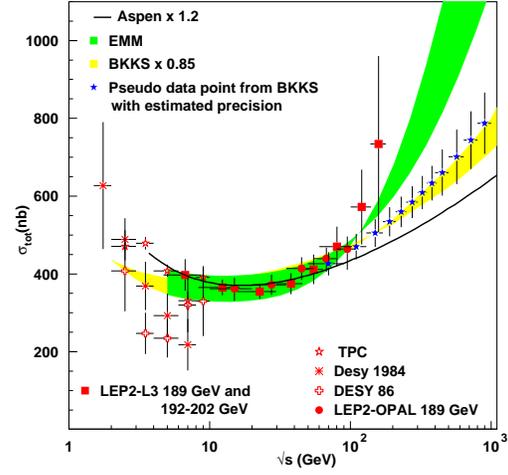}}
%
%
\caption{
The total \ggx \
cross-section as function of the \ggx \
collision energy, 
compared with model calculations~\cite{total}:
BKKS band (upper and lower limits correspond to different   
photon densities) and
EMM band (Eikonal Minijet Model for total and 
inelastic cross-section, with different photon densities and 
different minimum jet transverse momentum).
}
\label{fig:sigmatot}
\end{center}
\end{figure}

\section{Physics Topics}
Two-photon physics is not new. Most $e^+e^-$ colliders have or had 
a program of two-photon physics, by using the 
photons emitted from the lepton beams, which follow the 
well known WWA~\cite{wwa} energy dependence. The known disadvantage is 
the rapidly decreasing photon flux with photon energy: for collisions 
with a fractional energy $\sqrt{s_{\gamma\gamma}}/2E_e^{beam}$ larger
than 0.1 (0.5) the $\gamma\gamma$ luminosity is reduced by a factor 100 (10000)
with respect to the $e^+e^-$ luminosity. 
Hence the PC opens a new opportunity for truly {\it high} energy
two-photon physics, which is not limited to QCD but competes  in 
searches for new physics  and measurements
of Higgs  properties.

 The cross sections for charged particle pair productions are
considerably  larger in $\gamma\gamma$ collisions 
than for $e^+e^-$ collisions and decrease more slowly with energy.
Hence one can study new particles far from threshold with higher rate.
E.g.~$WW$ pair production in $\gamma\gamma$ at 500 GeV is a factor 20  
larger than
in $e^+e^-$. 
Cross sections for charged scalars, lepton and top pairs are a factor
$5-10$ higher at a PC, compensating for the reduced luminosity compared
to $e^+e^-$.

\subsection{QCD}
First we consider 
the QCD aspects of two-photon collisions in the reaction
$\gamma\gamma \rightarrow$ hadrons.
The nature of the photon is complex. A high energy photon can fluctuate
into   a fermion pair or into a bound state, i.e. a vector meson with
the
same quantum numbers as the photon $J^{PC} = 1^{--}$.
These  quantum fluctuations 
lead to the
so-called hadronic structure of the 
photon.

Many QCD studies of  photon-photon collisions were made for the 
TDR~\cite{pctdr}
and will not be repeated here. During this workshop we got new 
paramterizations 
of the energy dependence of the total cross 
section~\cite{pancheri,kwiecinski}, and new LO parametrizations
of the photon structure functions~\cite{jankowski}.

As an example   the  total \ggx \
cross-section is briefly discussed, a quantity that is
 not yet understood from first principles.
Fig.~\ref{fig:sigmatot} shows 
 the present photon-photon cross-section data
in comparison with recent phenomenological 
models~\cite{pancheri}.
All models predict a rise of the cross-section with the collision 
energy, \syy, but the amount of the rise
differs and 
predictions for high photon-photon  energies show noticable
differences. 
{\it Proton-like-models}
follow  closely the rise
 of the proton-proton cross-section, while in {\it QCD based} 
models, a stronger rise is 
predicted using the eikonalized pQCD jet cross-section. 

The figure demonstrates that large differences between the models 
become apparent in the energy  range  
of  a future 0.5-1 TeV \ep collider. 
An overview of new model
predictions is reported in~\cite{pancheri}.
The absolute precision with which
these cross-sections can be measured ranges from 5\% to 10\%,
where the largest contributions to the errors are due to 
the control of the diffractive component  of the cross-section,
Monte Carlo models used to correct for the event selections,
the absolute luminosity and knowledge on the 
shape of the luminosity spectrum~\cite{total}.
These prospects for measurement 
have been updated to the TeV range and are shown
in Fig.~\ref{fig:sigmatot}.

\subsection{Higgs Studies}

The quest for the Higgs particle(s) and the measurement of its properties 
will be 
one of the most important topics for high energy collider physics 
in the coming years. 
 The PC  is an ideal place to study the
 Higgs boson since    it can be produced as an 
 s-channel resonance.
 The mass reach of the PC is up to 80\% of the  
CMS energy of the 
 $e^-e^-$ collider. A detailed study of the $\gamma\gamma H$ vertex is only 
possible at a PC. Accurate measurements of mass and width are 
extremely important
and can be used to compare the SM predictions with those of
alternative models e.g. based on SUSY. Since the 
two-photon decay width of the Higgs  is sensitive 
to all heavy charged particles which acquire mass via the 
Higgs mechanism, the 
 partial
width could be modified  by 
 5-10\% in these models.

For a light Higgs, the 
  most promising channel is $\gamma\gamma\rightarrow H\rightarrow 
b\overline{b}$.
A first study
based on detector simulation,  showed that a 2\% statistical precision 
for the partial width could be reached~\cite{soldner}, for a Higgs
with mass of 120 GeV. During 
this 
workshop we have
\begin{itemize}
\item Revisited the $H\rightarrow b\overline{b}$ channel in detail
\item Studied the $H\rightarrow WW,ZZ$ channels
\item Studied analysis methods for the spin and CP properties of the Higgs
\item Studied the model separation power 
\item Studied the MSSM higgs
\end{itemize}
Members of the US PC study group have been reporting to us on their 
Higgs analyses as well, in particular $H,A$ production and 
discovery, the $H\rightarrow 
\gamma\gamma$ decay mode, and charged Higgs studies.

\begin{figure}[htb]
\centering
 \epsfig{file=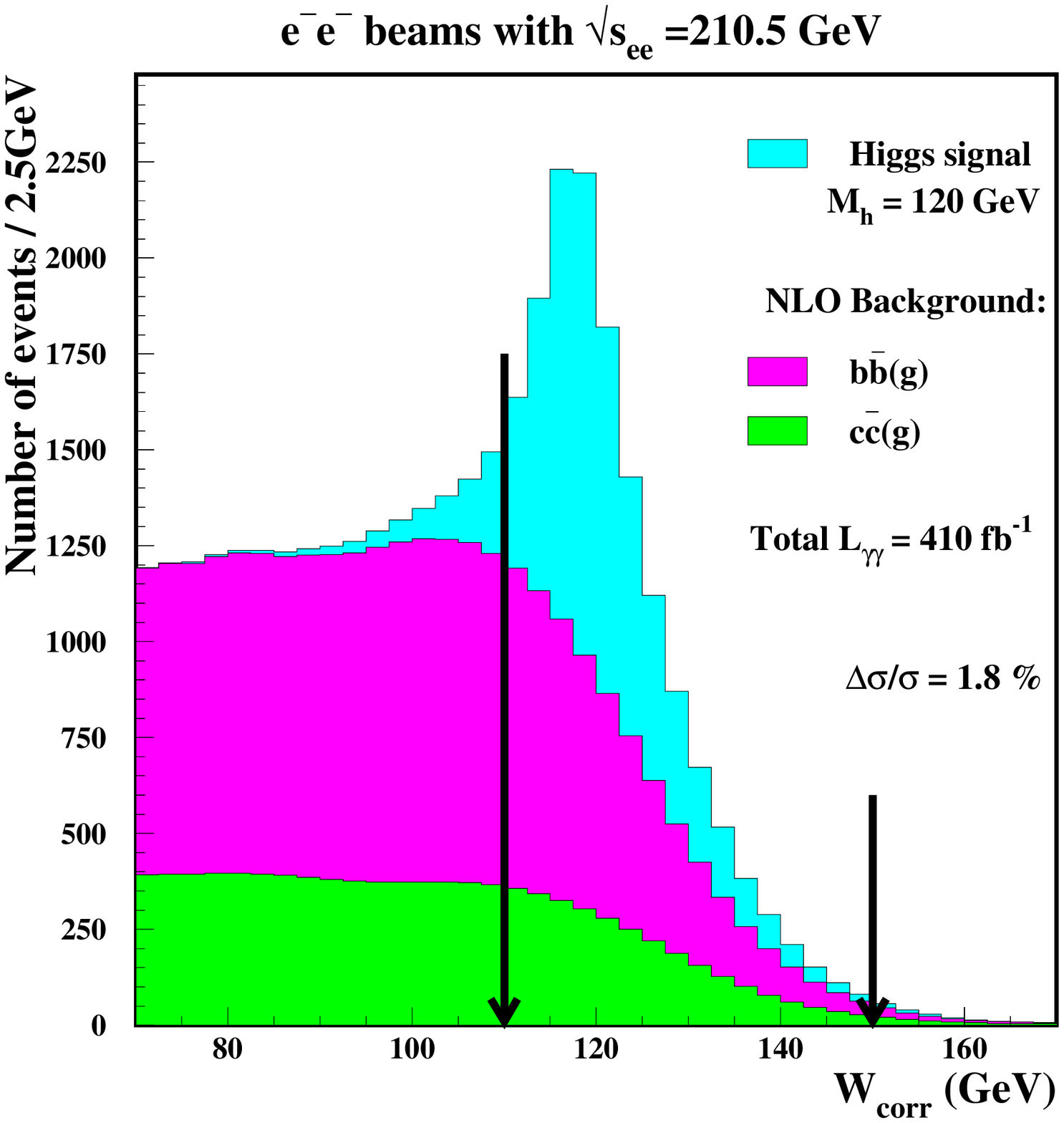,bbllx=0pt,bblly=0pt,bburx=540pt,bbury=570pt,height=6.5cm}
 \epsfig{file=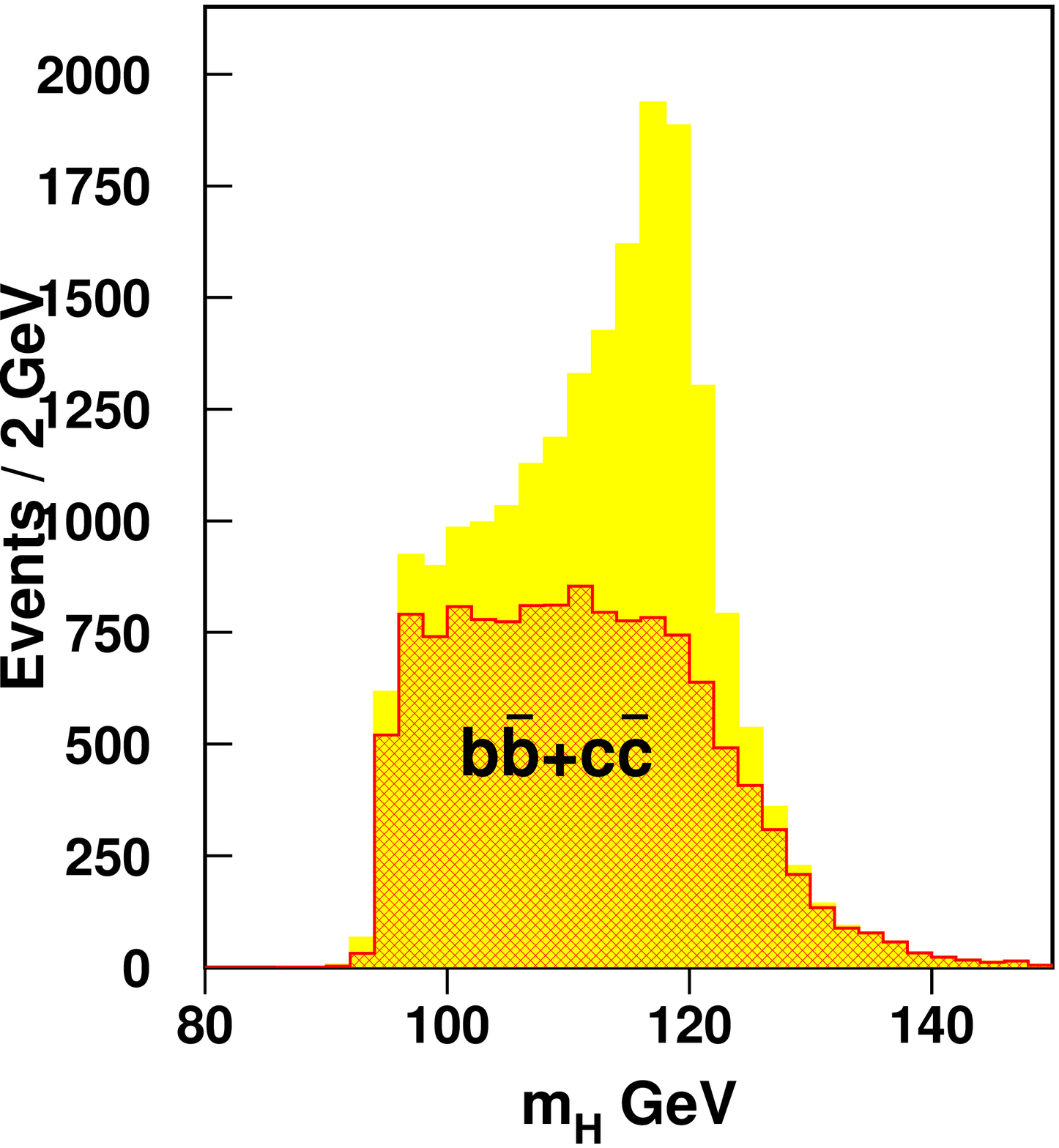,bbllx=20pt,bblly=0pt,bburx=566pt,bbury=620pt,height=6.5cm}
\caption{Reconstructed invariant mass distribution, 
$W_{rec}$ (top~\cite{piotr}) and $M_H$ 
(below\cite{rosca}), for selected
$b\overline{b}$ events. Contributions for background and signal are shown 
separately. In the top plot the 
arrows show the optimized mass window for the partial width
measurement.}
\label{fig:higgs1}
\end{figure}

\begin{figure}[htb]
\centering
 \epsfig{file=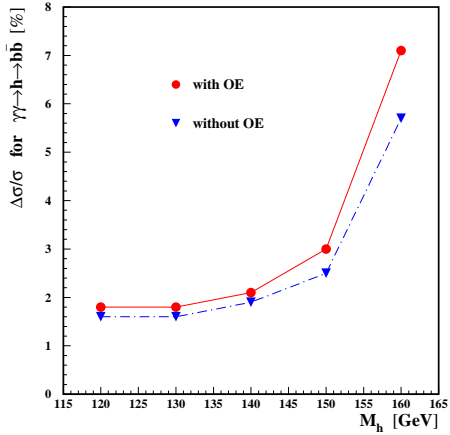,bbllx=0pt,bblly=0pt,bburx=500pt,bbury=530pt,height=6.0cm}
\caption{Statistical precision of 
$\Gamma(h\rightarrow \gamma\gamma)Br(h\rightarrow b\overline{b}$) measurements
for the SM Higgs boson with mass 120-160 GeV, with and without overlaying 
events (OE). The lines are drawn to guide the eye~\cite{piotr}.}
\label{higgsbbsig}
\end{figure}

First we discuss the $H\rightarrow b\overline{b}$ studies.
Selecting $J_z=0$ strongly suppresses the (Leading Order) contributions
of $b\overline{b}$ and 
$c\overline{c}$ production, but a good tagging of bottom quarks
with simultaneous rejection of charm quarks is needed.
During this workshop two new complete analyses were 
finalized~\cite{piotr,rosca}. The two studies use a different approach
for the background process, but come to the same conclusions.
The simulated mass spectrum
 for a Higgs particle with mass of 120 GeV, is shown in 
Fig.~\ref{fig:higgs1} for signal and background.
The PC will determine the quantity 
 $\Gamma(H\rightarrow\gamma\gamma)\cdot BR(H\rightarrow b\overline{b})$.
A feasibility study for a light Higgs, using a parametrized simulation
of the detector, has confirmed that 
the quantity above can be determined with a typical statistical
accuracy of about 2-3\%, 
as shown in Fig.~\ref{higgsbbsig}. These studies use as before the NLO
QCD backgrounds~\cite{jikia}.
New in these studies are the use of a more realistic photon spectrum, 
inclusion of overlap background QCD events (on average one event per 
bunch crossing), $b$-tagging using a neural net, and using a correction 
method for the reconstructed Higgs mass, accounting for  escaping 
neutrinos from the heavy flavour 
decays.


\begin{figure}[htb]
\centering
 \epsfig{file=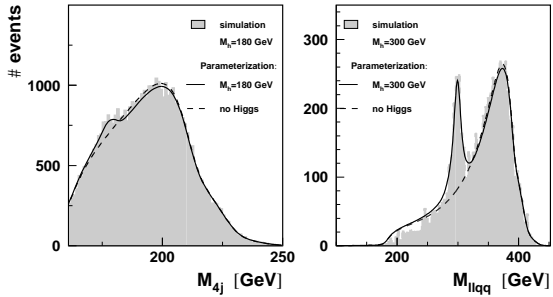,bbllx=0pt,bblly=0pt,bburx=660pt,bbury=400pt,height=5.0cm}
\caption{ Distribution of the reconstructed invariant mass 
$\gamma\gamma \rightarrow WW,ZZ$ for Higgs mass/electron beam of 
180/152.5 GeV and 300/250 GeV respectively~\cite{zarnecki}.}
\label{fig:wwzz}
\end{figure}

Since the $b\overline{b} $ branching ratio can be measured at an
$e^+e^-$ collider with a precision of 1-2\%,
 $\Gamma(H\rightarrow\gamma\gamma)$ can
 be determined with a statistical accuracy of approx. 2\%  for 
 an integrated luminosity of $85$ fb$^{-1}$, i.e one year running.

\begin{figure}[htb]
\centering
 \epsfig{file=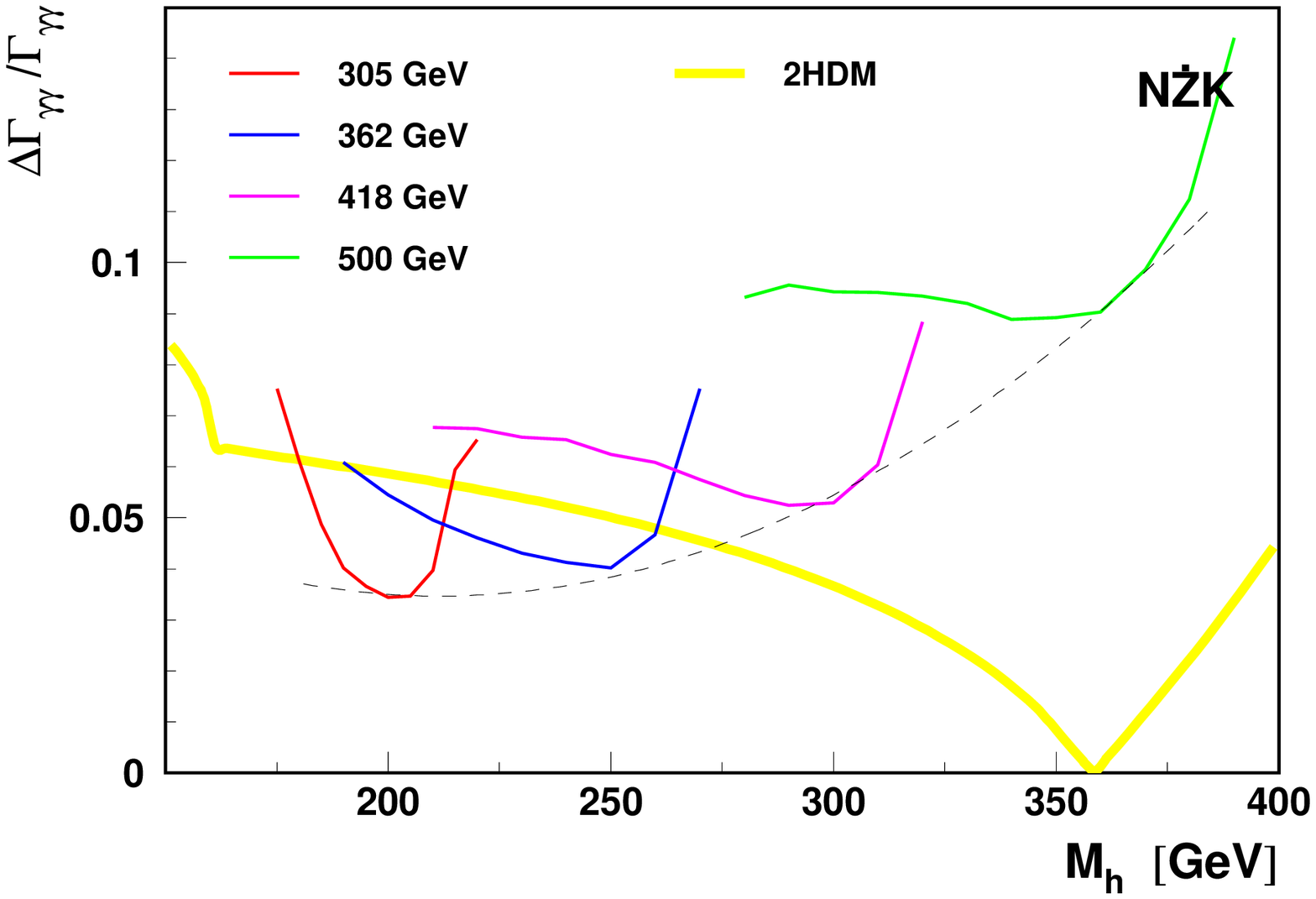,bbllx=0pt,bblly=0pt,bburx=550pt,bbury=400pt,height=4.5cm}
 \epsfig{file=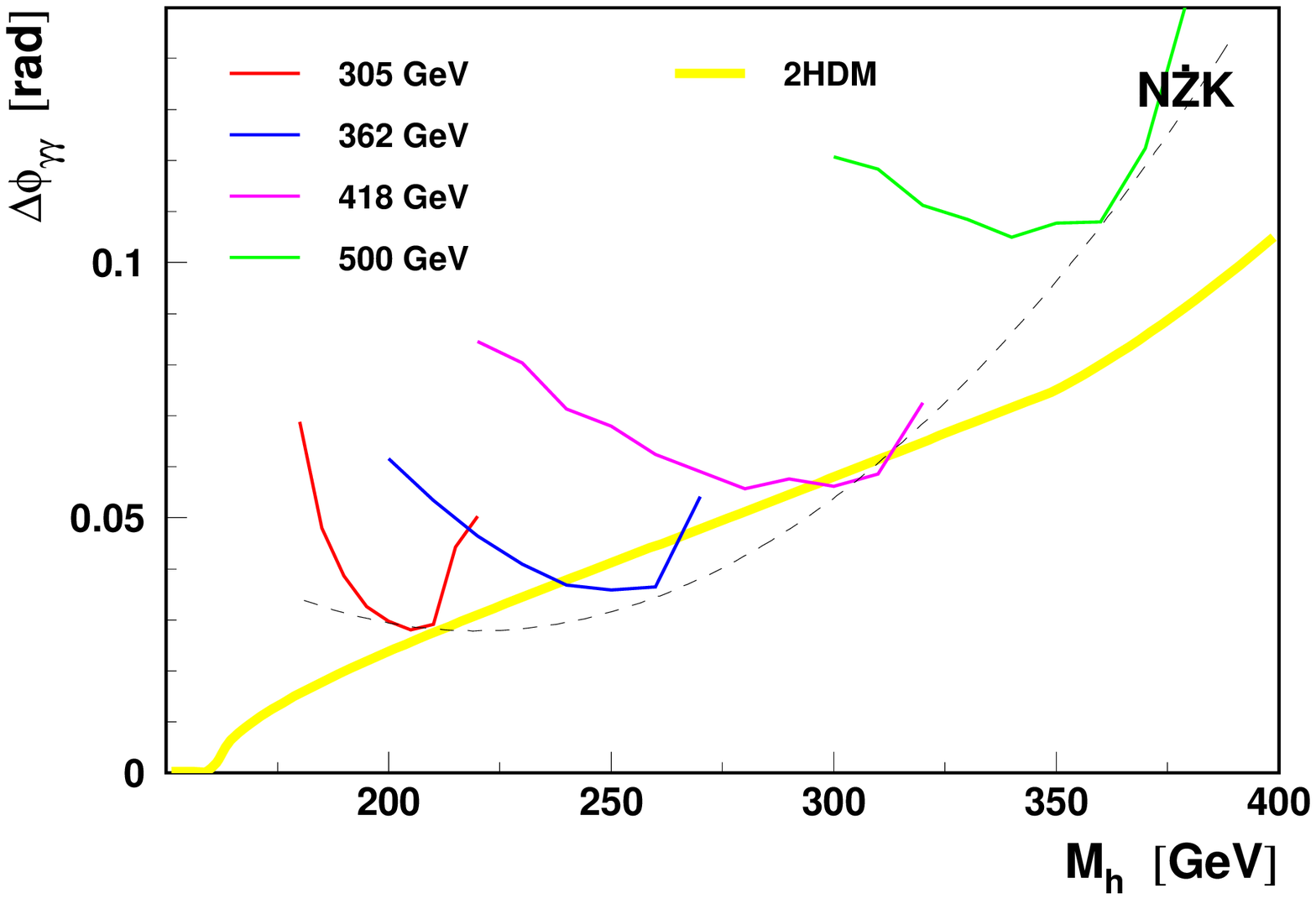,bbllx=0pt,bblly=0pt,bburx=550pt,bbury=400pt,height=4.5cm}
\caption{ Statistical determination of the Higgs boson width (upper plot)
and phase (lower plot) from the combined fit to the observed 
$WW$ and $ZZ$ mass spectra as a function of $M_H$~\cite{zarnecki}. 
The yellow (thick) 
line shows
the size of the deviation expected in the SM-like 2HDM II~\cite{mariax},
with an additional charged Higgs of 800 GeV.
The dashed line is to guide the eye.}
\label{fig:wwhiggs}
\end{figure}


\begin{figure}[htb]
\centering
 \epsfig{file=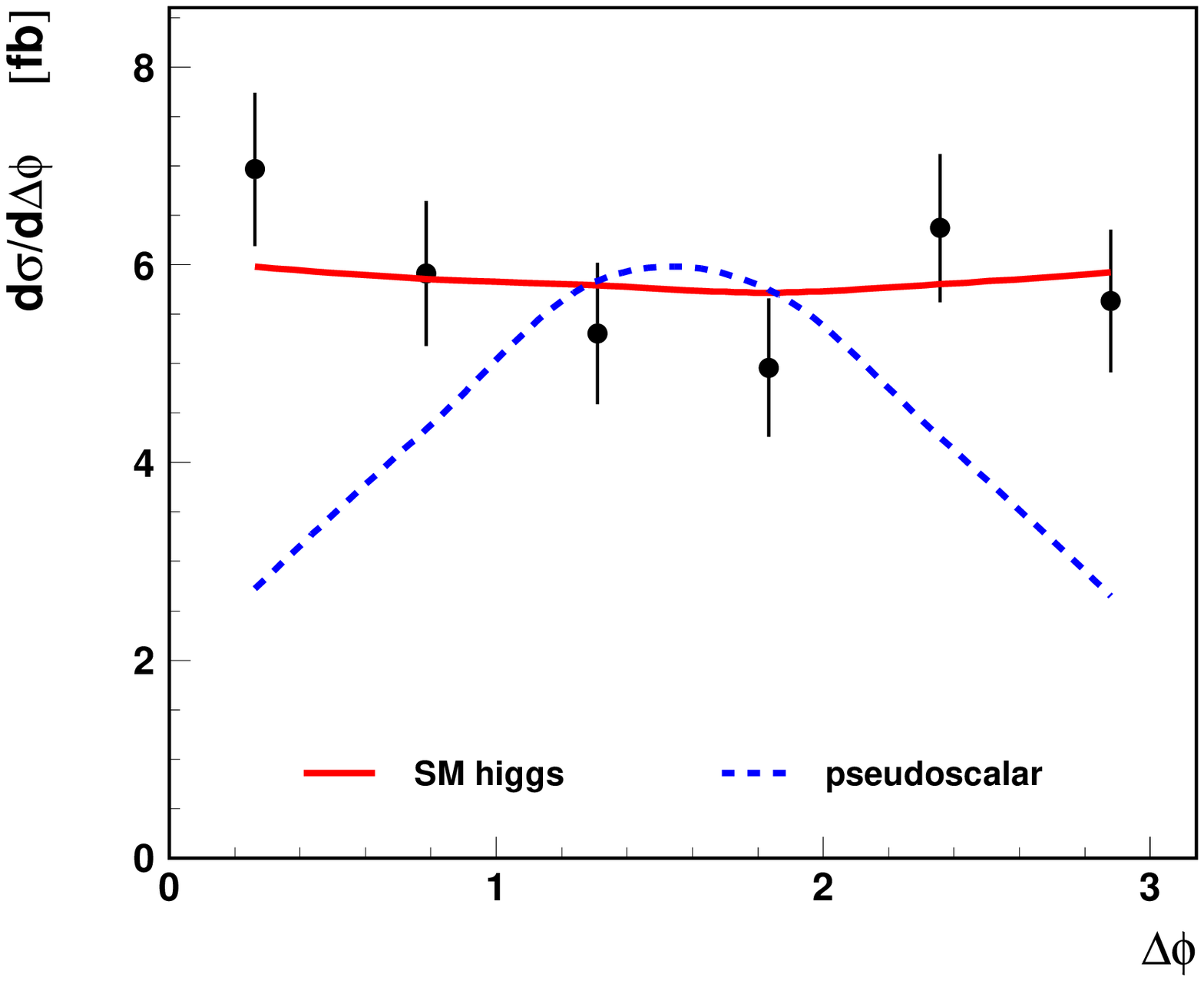,bbllx=0pt,bblly=0pt,bburx=500pt,bbury=400pt,height=4.5cm}
\epsfig{file=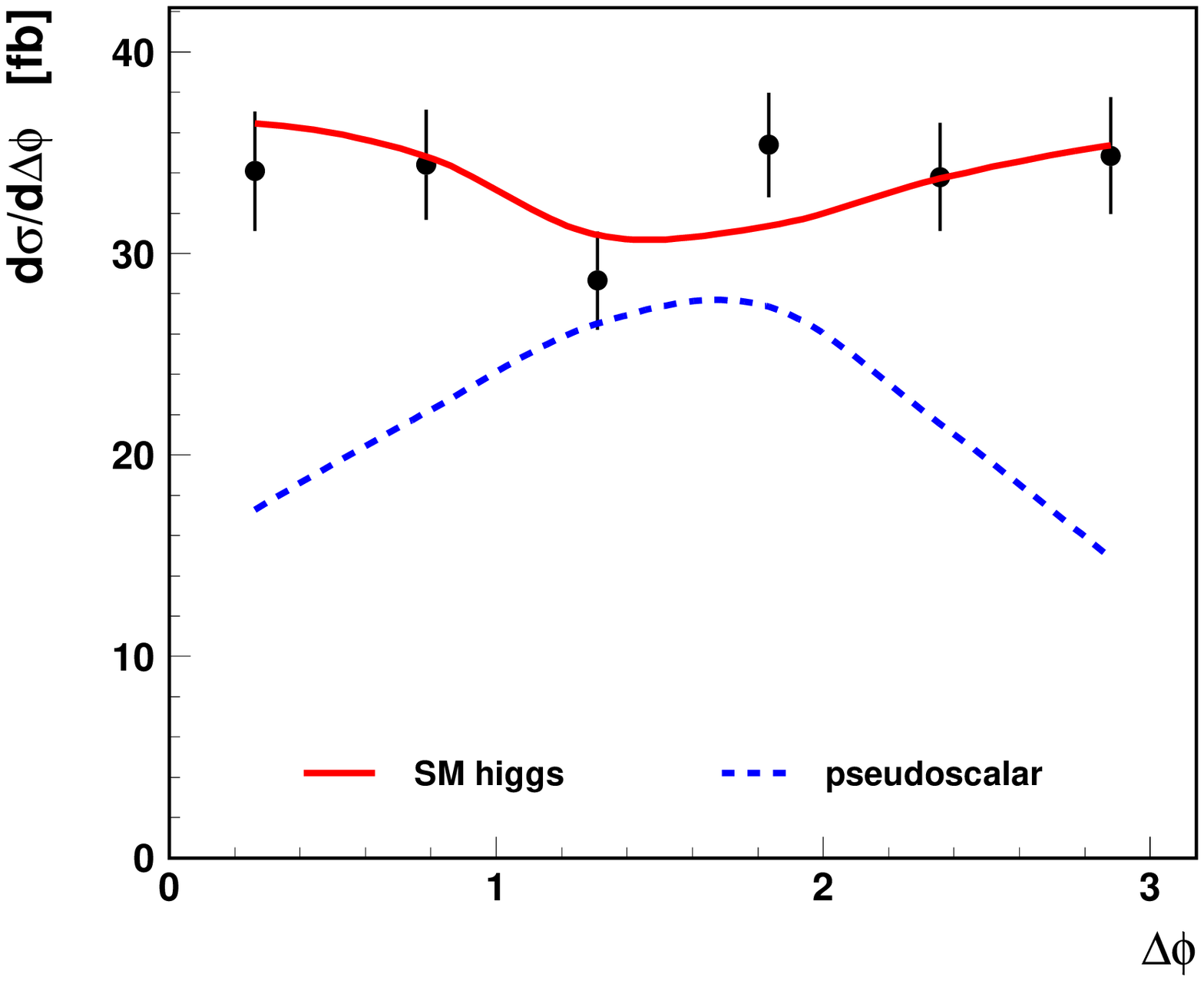,bbllx=0pt,bblly=0pt,bburx=500pt,bbury=400pt,height=4.5cm}
\caption{Statistical error on the determination of the
azimuthal angle $\Delta\phi$, as explained in the text, for 
$H\rightarrow WW$ and $H\rightarrow ZZ$.}
\label{ww-higgs3}
\end{figure} 

\begin{figure}[htb]
\centering
 \epsfig{file=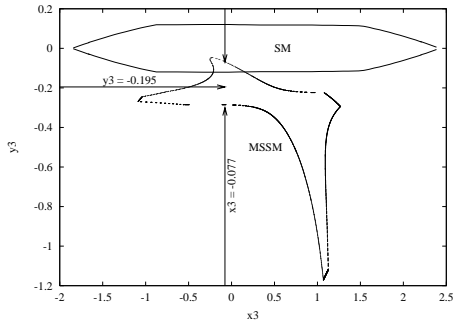,bbllx=0pt,bblly=0pt,bburx=450pt,bbury=350pt,height=6.0cm}
\caption{The boundaries of blind regions in the parameter space at 
95\% C.L. in the $x_3-y_3$ plane for a luminosity of 100 fb$^{-1}$ for 
$E_B$ = 310 GeV, given for both the SM and an example of a MSSM point.
Details on this analysis and definitions of the $x_3,y_3$
variables can be found in~\cite{godbole}. }
\label{fig:godbole}
\end{figure}

\begin{figure}[htb]
\centering
 \epsfig{file=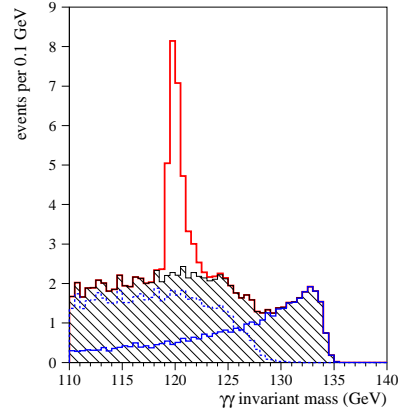,bbllx=0pt,bblly=125pt,bburx=550pt,bbury=630pt,height=6.0cm}
\caption{The  $\gamma\gamma$ invariant 
mass distribution including backgrounds from 
$\gamma\gamma \rightarrow \gamma\gamma$ (dashed lines)
and $e\gamma \rightarrow e\gamma$ (heavy solid line) as 
well as the Higgs signal peak~\cite{asner2}.
The hatched histogram shows the sum of the background contributions.}
\label{fig:higgsgamma}
\end{figure} 


\begin{figure}[htb]
\centering
 \epsfig{file=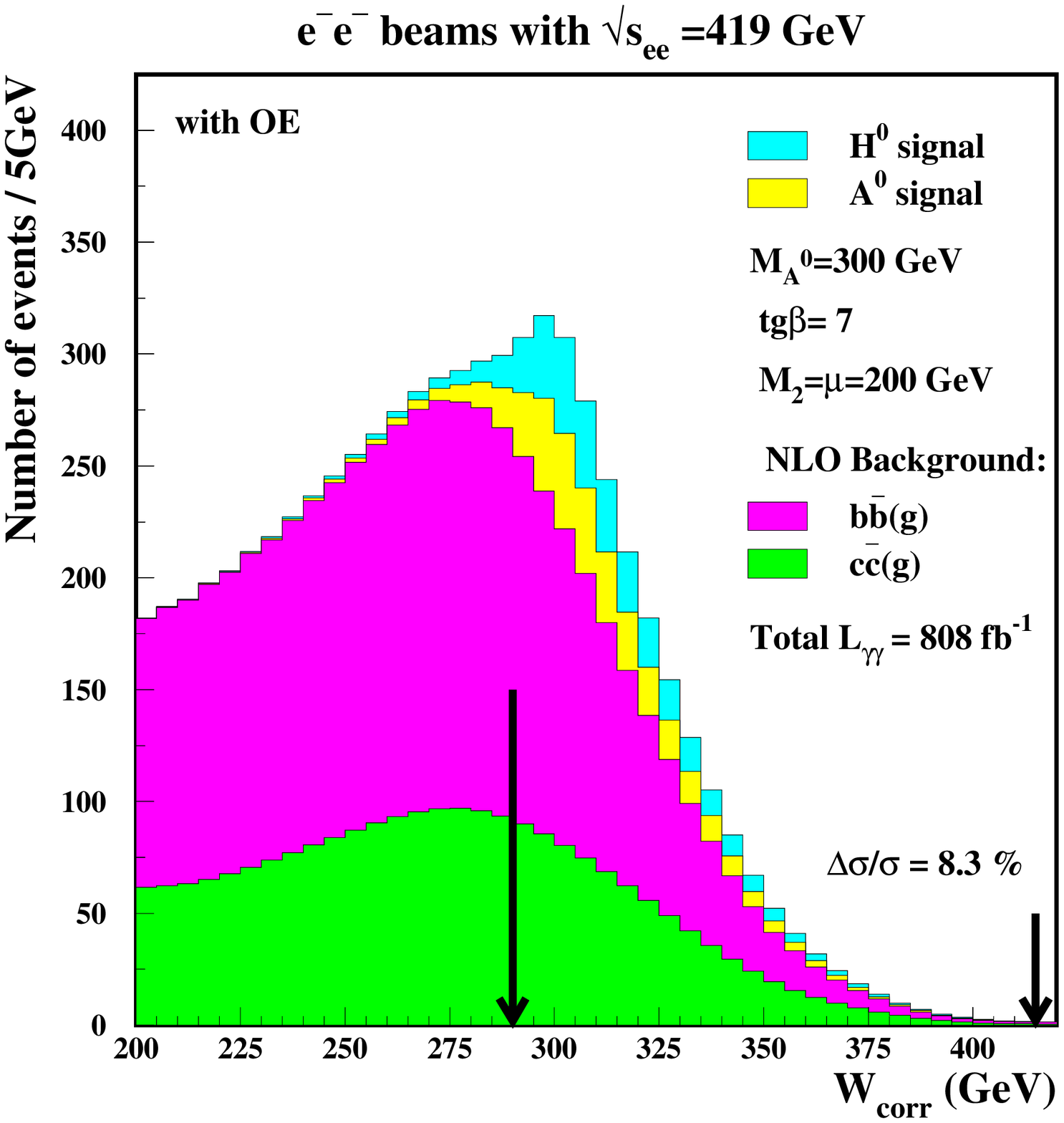,bbllx=0pt,bblly=0pt,bburx=540pt,bbury=570pt,height=6.5cm}
\caption{Reconstructed invariant mass, $W_{rec}$, distribution for selected
$b\overline{b}$ events for $H,A$.
 Contributions for background and signal are shown 
separately. The arrows show the optimized mass window for the partial width
measurement~\cite{zarnecki2}.}
\label{fig:HAmass}
\end{figure}

In \cite{zarnecki} the processes
 $\gamma\gamma \rightarrow H \rightarrow WW$ and
 $\gamma\gamma \rightarrow H \rightarrow ZZ$
have been studied for the region 180 GeV $<m_H< 350$ GeV 
 via  $q\overline{q}q\overline{q}$ decays
for the $WW$ channel  and  $llq\overline{q}$ decays  
for the $ZZ$ channel.
Typical mass plots are shown in Fig.~\ref{fig:wwzz}.
Due to the interference with the standard model background the processes
$\gamma\gamma \rightarrow$ Higgs $\rightarrow WW/ZZ$ turn out to be also
sensitivity to the phase of the $\gamma\gamma \rightarrow $ Higgs coupling,
$\phi_{\gamma\gamma}$. The measurement of both the phase and 
partial width gives powerful tools to discriminate a SM Higgs from
that of an extended model. 
A plot showing the sensitivity that can be reached on the partial 
two-photon width and the phase versus the mass of the Higgs is given in 
Fig.~\ref{fig:wwhiggs}, using the same 
simulation tools as for the light Higgs discussed above. 
Over a large region a sensitivity of 3-5\%
can be achieved.
The deviation from the SM prediction expected by a Higgs in a 
2HDM is also indicated~\cite{zarnecki}

Furthermore the CP structure of the Higgs boson can be verified
by studying the decay into $ZZ,WW$ and measuring the azimuthal 
angle $\Delta\phi$
between the decay planes of the two $Z,W$ bosons.
An example of the sensitivity of the 
angle $\Delta\phi$ is shown in Fig.~\ref{ww-higgs3}
for the decay channels $H\rightarrow ZZ,WW$, using a realistic simulation
and for one year of data taking.
In~\cite{zarnecki} one can find a very  extensive discussion on 
 sensitivities to CP properties using this and other variables, showing 
that a PC is an excellent tool for such analyses.

Further interesting CP studies include the study of 
 the channel
$\gamma\gamma \rightarrow t\overline{t}$, measuring 
asymmetries composed of the initial lepton beam polarization 
and the decay lepton charge~\cite{godbole}.
A sensitivity plot is shown in Fig.~\ref{fig:godbole}.

\begin{figure}[htb]
\centering

\epsfig{file=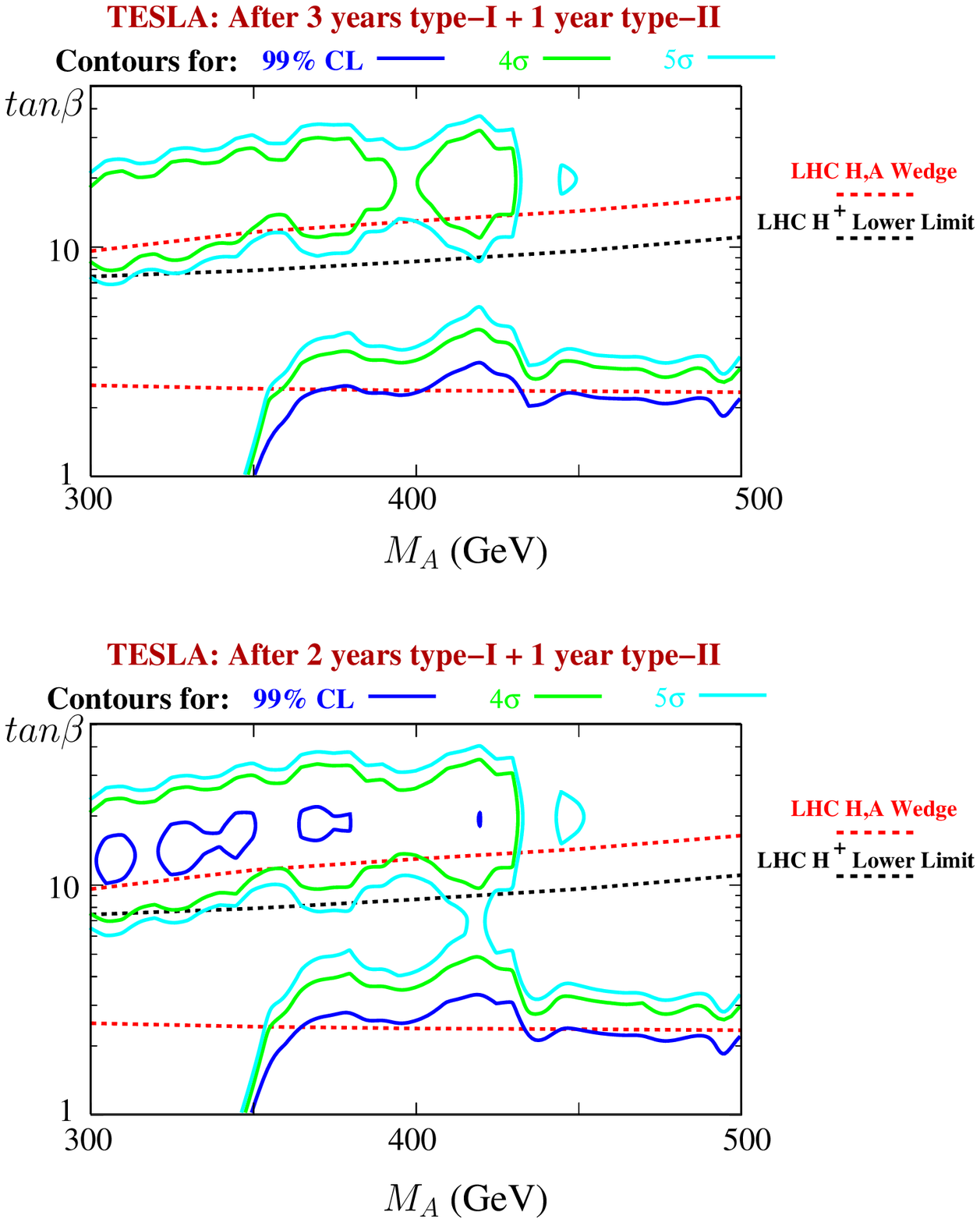,bbllx=0pt,bblly=130pt,bburx=490pt,bbury=700pt,height=8.0cm}
\caption{ Regions in $M_A$, $\tan\beta$ where the LHC has problems 
discovering the heavy Higgs, $A$ and $H$, with the statistical 
sensitivity at a PC based on a 630 GeV $e^+e^-$ collider,
after several years of running~\cite{asner}.}
\label{fig:HA}
\end{figure} 

\begin{figure}[htb]
\centering
 \epsfig{file=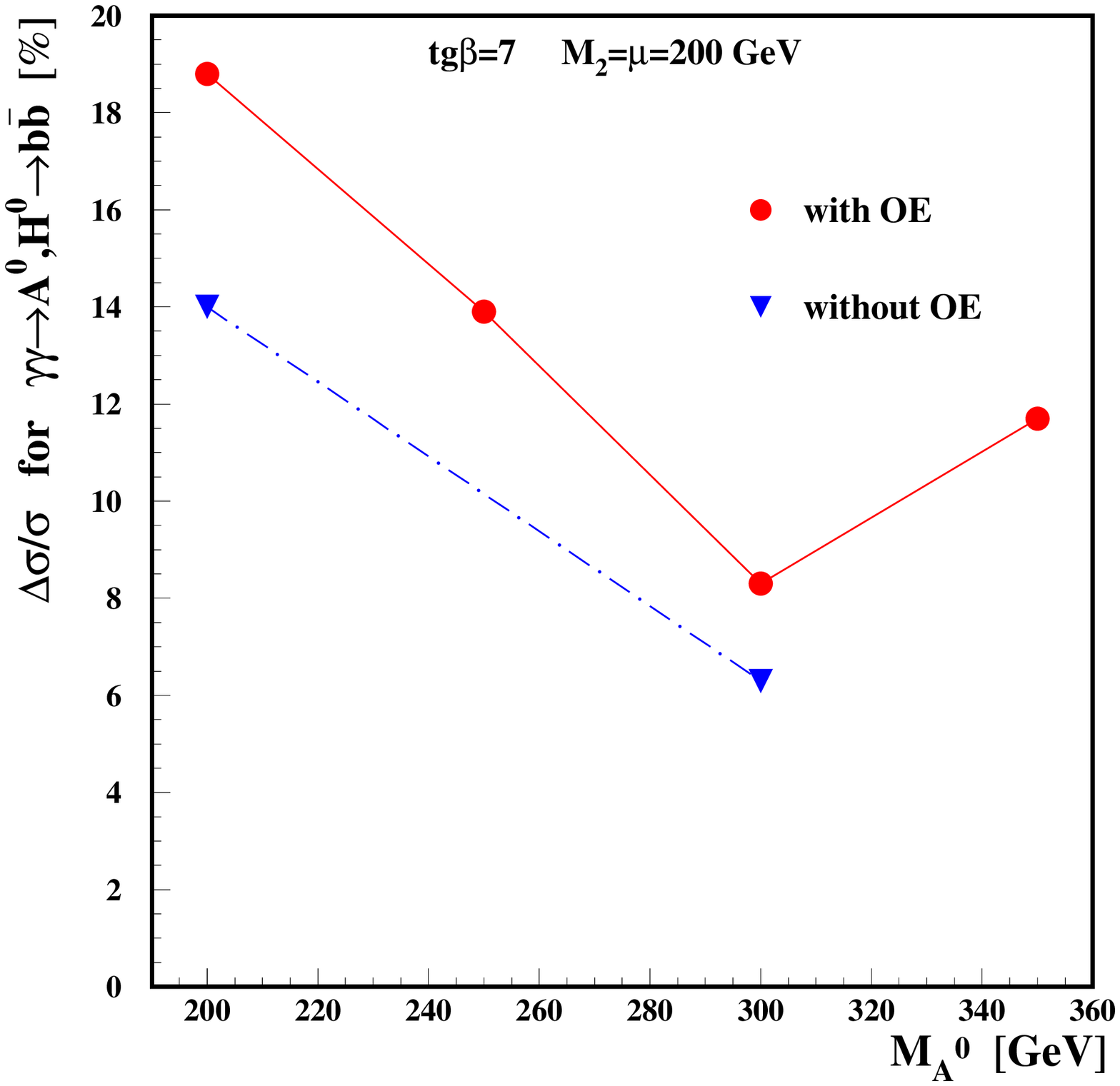,bbllx=0pt,bblly=0pt,bburx=500pt,bbury=530pt,height=6.0cm}
\caption{Statistical error on the determination of the
$\sigma(\gamma\gamma \rightarrow A,H \rightarrow b\overline{b}$)
measurements shown for $M_A = 200-350$ GeV and $\tan\beta=7, M_2=\mu=200$
GeV, with and without overlaying events (OE)~\cite{zarnecki2}. 
The lines are to guide the 
eye.}
\label{fig:HAsignif}
\end{figure}

Our US colleagues have reported to us on studies of 
 $\gamma\gamma \rightarrow H \rightarrow 
\gamma\gamma$ and charged Higgs production.
The first channel is quadratically sensitive to the two photon
Higgs partial width. The event rate is however small and  an 
excellent 
calorimeter is need for the signal to be observable. In the analysis 
a calorimeter energy resolution
$\sigma_e/E=((0.015/\sqrt{E})^2 +(0.0045)^2)^{1/2}$ 
was assumed which is better
than the CMS experiment EM  calorimeter resolution. 
This would be also a different calorimeter than what is currently
envisaged for the TESLA detector.
The signal for one year of running 
is shown in Fig.~\ref{fig:higgsgamma}.
The mass resolution on the peak is 0.4 GeV, allowing for a measurement of
$\Delta m_H \sim 100$ MeV and  $\Delta\sigma/\sigma$ of 24\%.
A crucial issue will be  the understanding of the background.

An analysis of the production of charged Higgses,
 which  appear in extended Higgs doublet models, is reported 
in~\cite{asner2}.
The cross section is about a factor 20 larger than for $e^+e^-$
collisions.
Taken into account the branching ratios, for a charged Higgs below
200 GeV generally the channel 
$\gamma\gamma \rightarrow H^+H^-  \rightarrow \tau \nu\tau\nu$
is the most promising. With suitable cuts (albeit with a very
low efficiency of a few \%) a S/B of about 3 can be achieved. This decay 
mode does not allow to reconstruct 
the mass. To get mass information the channel  
$H^+H^-  \rightarrow \tau \nu q\overline{q}$
is under study.
More PC studies of the US group are reported in ~\cite{asner2,asner1}.

An important ``golden'' 
channel for the PC is the production $\gamma\gamma \rightarrow
H,A$. Indeed, a PC may help to discover $H,A$ bosons in the MSSM
SUSY extension of the SM  when these are inaccessible
by other machines.
For example the LHC cannot extract the $H,A$ signals out of the background
(except perhaps for SUSY decay modes of the $H,A$) if the mass is larger
than about 200-300 GeV at medium $\tan\beta$.
Fig.~\ref{fig:HAmass} shows the mass distribution of the $H,A$ in the 
$b\overline{b}$ decay channel. This mass distribution~\cite{zarnecki2} 
was  estimated
using exactly the same tools as for the light Higgs $H\rightarrow 
b\overline{b}$ analysis~\cite{piotr}.
Fig.~\ref{fig:HA} shows the region that could be covered by a PC
for several years of running (assuming a 630 GeV collider)~\cite{asner}
in the $b\overline{b}$ decay mode. The $e^+e^-$ mode of that collider 
can reach $M_{H,A}$ masses up to about 300 GeV only. 
The PC essentially closes the wedge left by the LHC, up to masses
of 500 GeV.
Fig.~\ref{fig:HAsignif} shows the precision with which the cross section can 
measured for $M_A$ in the range of 200-350 GeV and $\tan\beta=7$, with and
without overlaying events.

\subsection{Standard Model}
Due to the large cross sections, several precise measurements of 
SM parameters or particle properties can be  made at a PC.

Triple gauge couplings were studied in detail, 
using realistic luminosity spectra and detector 
simulation~\cite{moenig_tgc}.
The WHIZARD~\cite{whizard} Monte Carlo was used for the signal.
The study shows that these couplings can be measured 
at a PC with a precision similar to the
one achieved at an $e^+e^-$ collider, see  Table~\ref{tabtriple}.
\begin{table}
\footnotesize{
\begin{tabular}{|c|c|c|c|}
\hline
(GeV)  & $\sqrt{s_{e\gamma}}$=  450  & $\sqrt{s_{\gamma\gamma}}$=
400  & $\sqrt{s_{ee}}$= 500  \\
  $\int L\Delta t $ &$ 110 fb^{-1}$ 
 & $ 110 fb^{-1}$ 
 & $ 500 fb^{-1}$ \\
$\Delta L$ & 0.1\% & 0.1 \% & \\
\hline
$\Delta K_{\gamma} $ &$ 9.9\cdot 10^{-4}$ & $6.7\cdot 10^{-4}$ & $3.1\cdot 
10^{-4}$ \\
$\Delta \Lambda_{\gamma}$ &$ 2.6 \cdot 10^{-4}$& $(6.0)\cdot 10^{-4}$ &$ 4.3\cdot 10^{-4}$ \\
\hline
\end{tabular}
}
\caption{Precision achievable on triple gauge couplings for a 
$\gamma\gamma$, $e\gamma$ and $ee$ collider.}
\label{tabtriple}
\end{table}
The sensitivity is proportional to the momentum of the particles 
involved in the triple gauge boson vertex. 
The analysis~\cite{moenig_tgc} includes detector 
simulation and 3D fits including the azimuthal decay angle (not yet
done for the
$\gamma\gamma$ study).

Top quark production was studied in~\cite{golden}. 
The $e\gamma$ scattering gives a good sensitivity to the 
anomalous top couplings, as detailed in that report.
The reaction $\gamma\gamma \rightarrow t\overline{t}$ allows for 
an extraction of the electric dipole moment: for
20 fb$^{-1}$ and an electron beam energy of 250 GeV a sensitivity on the 
dipole moment of $1.3\cdot 10^{-16}$ ecm can be achieved, when assuming
a realistic luminosity spectrum~\cite{godbole_dip}.

\begin{figure}[htb]
\centering
 \epsfig{file=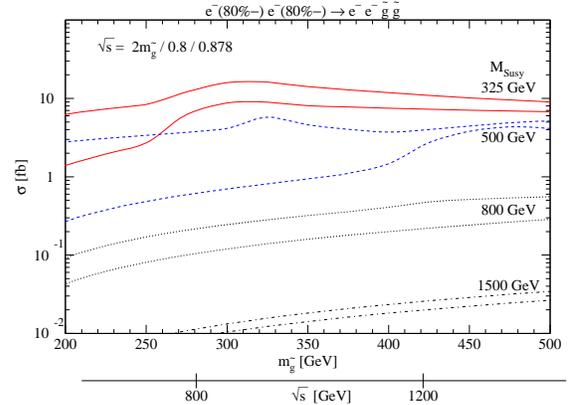,bbllx=0pt,bblly=0pt,bburx=550pt,bbury=450pt,height=6.0cm}
\caption{ Cross section for gluino production in $\gamma\gamma$ collisions
versus the gluino mass and for different squark masses~\cite{berge}
(maximal stop mixing (thin lines) and no mixing (thick lines)).}
\label{fig:gluinos}
\end{figure} 


\begin{figure}[htb]
\centering
 \epsfig{file=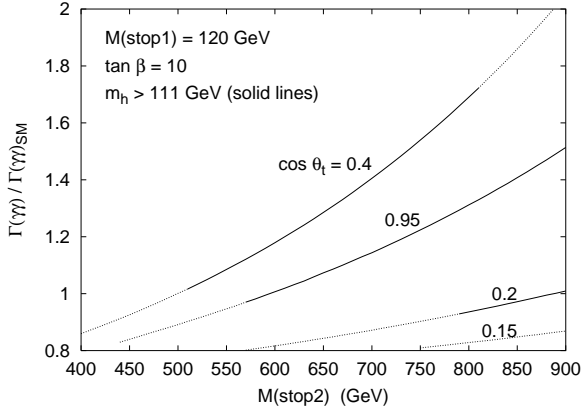,bbllx=-350pt,bblly=50pt,bburx=400pt,bbury=650pt,height=5.5cm}
\caption{Dependence of the partial two photon width of the Higgs  on 
$M_{\tilde{t}_2}$
for various values of $M_{\tilde{t}_1}$. Here $M_A$ is 1 TeV, 
$\tan\beta =10 $ and $M_2=-\mu = 200$ GeV, other SUSY mass parameters
are set to 1 TeV~\cite{asner2}.}
\label{fig:stop}
\end{figure}

\begin{figure}[htb]
\centering
\epsfig{file=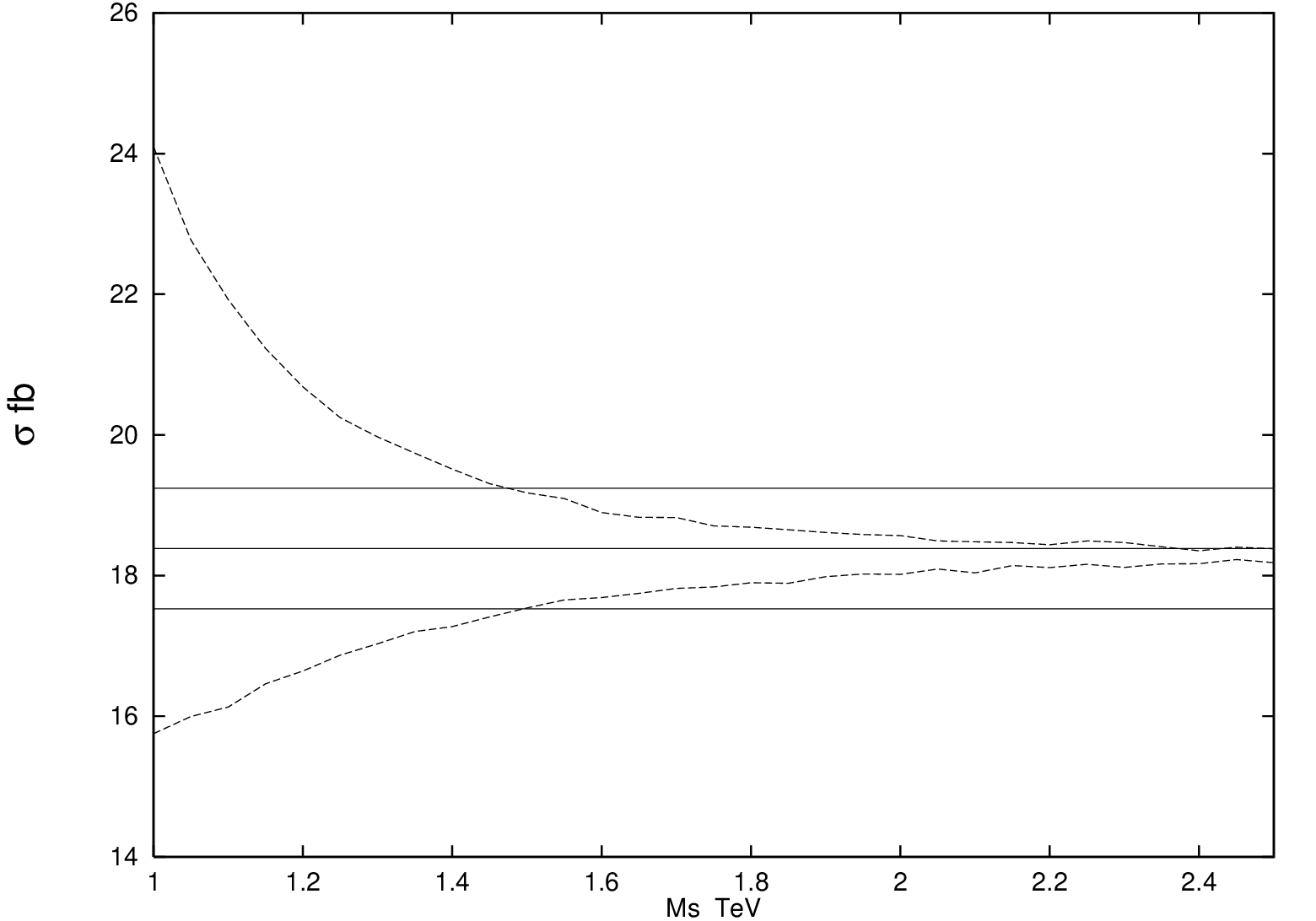,bbllx=100pt,bblly=60pt,bburx=650pt,bbury=600pt,height=4.0cm}

\epsfig{file=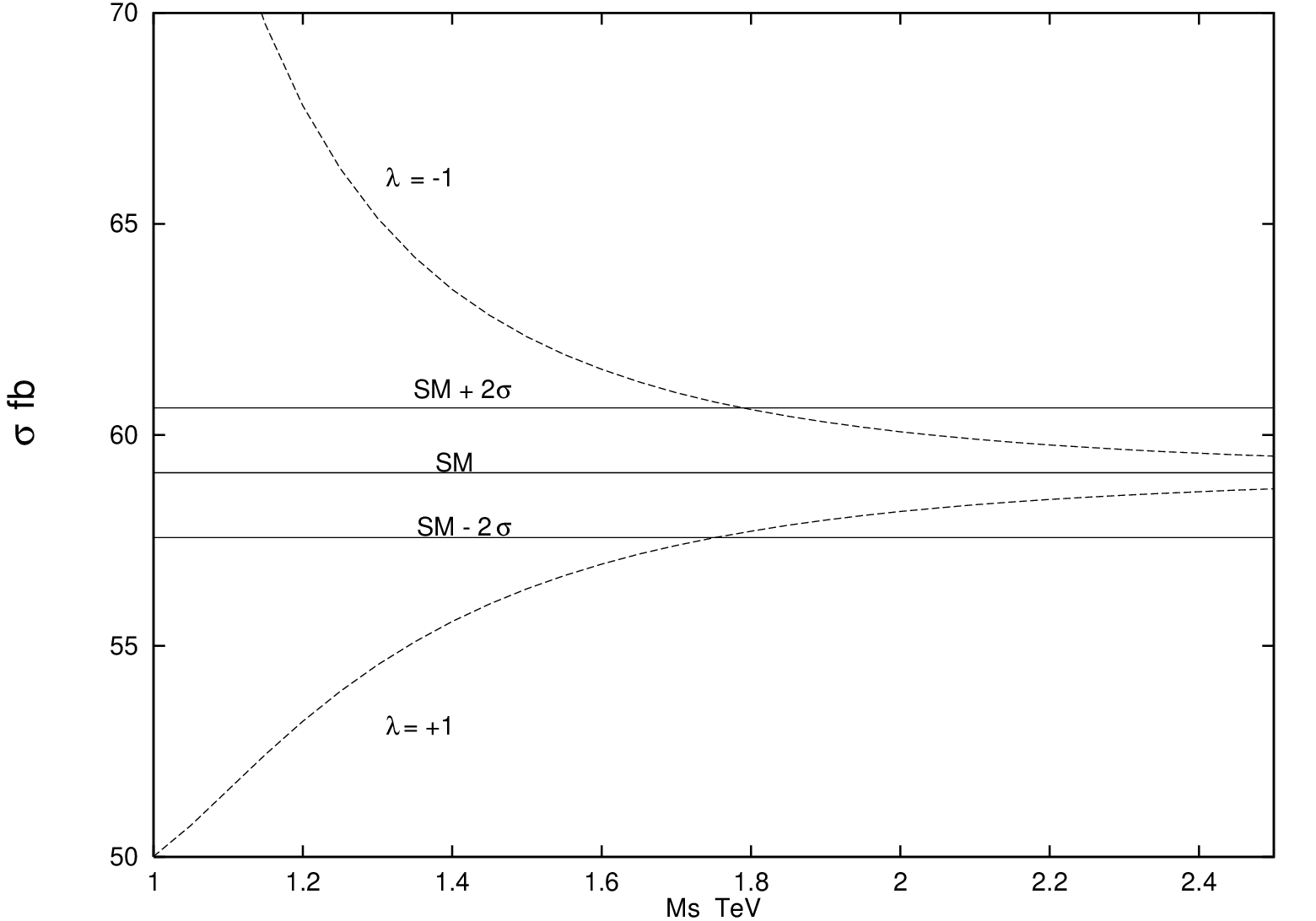,bbllx=100pt,bblly=60pt,bburx=650pt,bbury=600pt,height=4.0cm}
\caption{The sensitivity to ADD extra dimensions in the channel
$\gamma\gamma \rightarrow t\overline{t}$, for an ideal Compton spectrum 
(top) and for a realistic one using CompAZ (bottom). }
\label{fig:addx}
\end{figure}

\subsection{Beyond the Standard Model}
Supersymmetry is presently the most popular theory for physics beyond the 
standard model. A few examples are given where a PC can make 
significant contributions.

If the LSP is light, the process 
$e\gamma \rightarrow \tilde{e}\chi^0_1 \rightarrow e \chi^0_1 \chi_1^0$
can extend the range of discovery for heavy sleptons. Indeed LHC has 
difficulties discovering sleptons for masses above 300-350 GeV, and the
$e^+e^-$ collider has to pair produce sleptons, hence its range
is  limited to 
$\sqrt{s}_{ee}/2$.
In case of a $e\gamma$ collider the 
reach is $0.9 \cdot \sqrt{s}_{ee} - m_{\chi^0_1}$, e.g. 350 GeV for  
250 GeV electron beams and a LSP of 100 GeV~\cite{watanabe}.

Another  channel of interest at a PC is 
 $\gamma\gamma \rightarrow $ gluinos.
This reaction is only accessible at an $e^+e^-$ collider if the 
squarks are heavier than the gluinos and the decays $\tilde{q}
\rightarrow \tilde{g}q$ are open.
Photons couple to squarks and quarks and can produce gluinos 
via box diagrams. The yield is shown in Fig.~\ref{fig:gluinos}. 
Typically 2000 gluinos pairs 
can be produced/year for light quarks (325 GeV)~\cite{berge}.
It remains to be seen what one can learn more at a PC
than what is known from the 
LHC at that point.

Measuring the two photon width at a PC can also help to pin down 
masses of sparticles which cannot be 
directly produced at the $e^+e^-$ collider.
An example is shown in Fig.~\ref{fig:stop}, where we assume a scenario 
of large mass splitting between the $\tilde{t_1}$ and $\tilde{t_2}$.
If the $\tilde{t_1}$ mass and $\tilde{t}$ mixing angle 
are known from $e^+e^-$ 
studies then using a precise measurement of the two-photon partial 
width of the Higgs one can constrain 
the mass of $\tilde{t_2}$  as shown in the Figure~\cite{asner1}.

Other new theories propose the existence of extra dimensions. It appears 
that the reaction 
$\gamma\gamma \rightarrow WW$ is very sensitive to ADD type of 
effects~\cite{rizzo}.
The sensitivity scales with a CMS energy as 11 $\sqrt{s}$.
For $e^+e^-\rightarrow f\overline{f} $ the sensitivity is
 6.5 $\sqrt{s}$, and for the LHC using the process 
$pp \rightarrow jj$ it is 9 TeV for 100 fb$^{-1}$.
A new study shows the sensitivity to ADD extra dimensions in the channel 
$\gamma\gamma \rightarrow t\overline{t}$
in Fig.~\ref{fig:addx}~\cite{godbole_add}: the top figure 
takes the ideal Compton spectrum while the lower figure includes the 
luminosity via CompAZ. The sensitivity is reduced from 
$M_s = 1.7$ TeV to 1.4 TeV for one year of running.

\section{Technology for a PC}

A photon collider IP introduces new challenges: The laser part, the optics,
stability and control in the IP (to 1 nm), length control in case of a 
cavity, beam extraction line, etc.
Both the European and the US groups have an R\&D effort on the hardware
part. 

\begin{figure}[htb]
\centering
 \epsfig{file=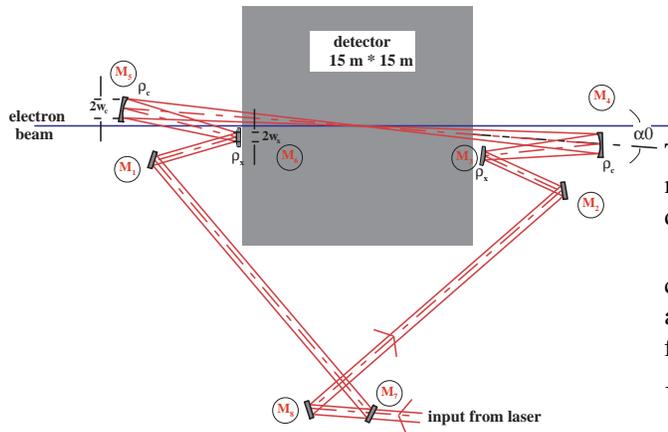,bbllx=0pt,bblly=0pt,bburx=1000pt,bbury=600pt,height=7.0cm}
\caption{The optical cavity for the TESLA PC IP design.}
\label{fig:cavity}
\end{figure}

Europe 
is developing a scheme for an optical cavity, shown in 
Fig.~\ref{fig:cavity}~\cite{berlin},
and plans are considered
to make a 1:9 scale model. The use of a cavity  
allows
multi-passing of the laser signal and thus reduces the required laser 
power.
The US group of LLNL follows a full power laser design, as the short 
bunch distance at the NLC is less favourable to benefit   
from  such a cavity option.

\begin{figure}[htb]
\centering
 \epsfig{file=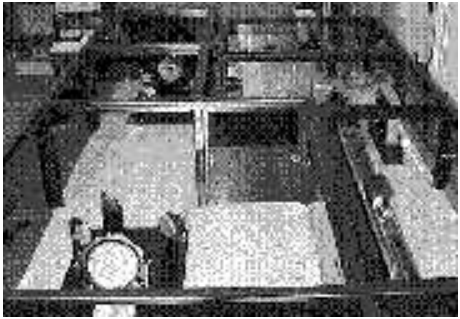,bbllx=0pt,bblly=500pt,bburx=175pt,bbury=618pt,width=0.7\linewidth}
\caption{The optics setup at LLNL.}
\label{fig:llnl}
\end{figure}

The US group has commissioned a laser with 20 J pulses at 10 Hz. The
full power (100 Hz at 10 Hz) 
is expected to be reached next year. In total 10 of these lasers
would be required. They have also studied interferometry for alignment,
built a half-size focusing optics setup in the lab, studied a beam-beam 
deflection feedback system, and are preparing a proposal for a PC testbed
at SLAC, using the SLC and perhaps even parts of the SLD~\cite{gronberg}.
A picture of the set-up of  the optics  is shown in Fig.~\ref{fig:llnl}.

\begin{figure}[htb]
\centering
\epsfig{file=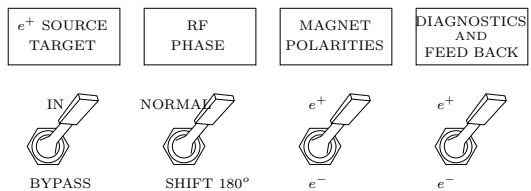,bbllx=150pt,bblly=600pt,bburx=490pt,bbury=750pt,width=9.0cm,clip=}
\caption{Since the $e^-e^-$ collider requires only minor changes to the
hardware of the $e^+e^-$ machine and detector, its programme could be 
pursued during the first face of the facility...'' International Linear 
Collider  Technical Review Committee Report- 1995~\cite{jejuheusch}.}
\label{fig:clem}
\end{figure}

In all there is progress but funding is presently certainly and issue
to continue the R\&D. The
developments during the  coming years will be of vital importance.

\section{e-e- collisions}
The PC will be based on  $e^-e^-$ collisions. These collisions can be 
of great interest by itself. 
No new studies have been presented in the context of this workshop, but 
an excellent overview paper can be found in the proceedings of 
the LCWS2002~\cite{jejuheusch}.
Here we recall on a few of the outstanding advantages of $e^-e^-$
\begin{itemize}
\item Large polarization for both beams, hence (almost) pure $e_L, e_R$ 
initial states.
\item Excellent discovery potential 
for states with exotic quantum numbers 
(e.g such as $H^{--}$)
\item Larger sensitvity (for identical luminosity) than $e^+e^-$
e.g. for contact interactions, non-commutative 
scales (via Moller scattering)
\item Special processes can be very clean, e.g. $e^-e^- \rightarrow 
e^-e^- H$
\item Sharper onset of e.g. the slepton production threshold than in $e^+e^-$
\item Possibility to identify TeV level Majorana neutrinos through the lepton 
number violation reaction $e^-e^- \rightarrow 
W^-W^-$
\end{itemize}
To be fully convincing these studies  need to reach the same maturity 
as for the $e^+e^-$ collider  or
 PC studies: i.e. include detector 
simulation, backgrounds, 
beamstrahlung, ...

On the downside there are of course no classical s-channel 
processes in $e^-e^-$, and since
the beams show an anti-pinch effect, the luminosity in general is 
lower than for $e^+e^-$. One finds typically numbers in the ball-park 
of $L_{e^-e^-} = 0.15-0.3 \cdot L_{e^+e^-}$~\cite{schreiber}.

Unlike for the PC there are however no major changes required in the 
interaction region or accelerator. The $e^-e^-$ option is the extra option
which for TESLA would be most easily to realize.
Fig.~\ref{fig:clem} shows how easy it could be for the machine shift
leader to switch from $e^+e^-$ to $e^-e^-$ collisions: just four
switches to turn...  
Clearly we must keep it on the roadmap and the future new physics 
will decide how valuable this option will be for us.

\section{Conclusion}

An $\gamma\gamma$ and $e\gamma$ collider will provide exciting 
 physics opportunities, many of which have been  developed in detail
during the last two years. The development of specific PC 
study tools has allowed that several of the studies have now reached the
necessary maturity.

At the LCWS2002 in Jeju a panel discussion was organized on the PC
option~\cite{gross}.
 The conclusion was a clear plea to continue the R\&D 
and 
physics studies such that 
we can be in a good position to incorporate a PC in the overall planning 
of a LC, when that day comes.
A PC will be largely complementary to its drive LC and will therefore
strengthen the case for such an $e^+e^-$ collider.
A PC option  should be 
considered from the onset within the planning of the project.
A vigorous R\&D plan for a PC will need to be put in place, preferably 
on a world-wide level.

Finally
an (updated) short list of processes which are considered to be  most 
important 
for the physics program of the photon collider option of the LC, is
presented in Table~2, taken from ref.~\cite{golden}.
Additionally to this list are the processes
$e\gamma \rightarrow e^*$, leptoquark production, strong WW scattering
and $e\gamma \to eH$.
It summarizes the rich physics program that becomes accessible at a
Photon Collider!

\begin{table}[!hbtp]

\label{processes}
\vspace{2mm}
\footnotesize{
\begin{center}
\begin{tabular}{ l  c } 
\hline
$\quad$ {\bf Reaction} & {\bf Remarks} \\
\hline\hline
$\GG\to  H,h\to b\bar b$ & SM/MSSM\  Higgs, 
 $M_{H,h}<160$~GeV \\
$\GG\to H\to WW(^*)$    & SM\ Higgs,
140$< M_{H}<190$~GeV \\
$\GG\to H \to ZZ(^*)$      & SM\ 
Higgs,  180$< M_{H}<350$~GeV \\
$\GG\to H \to \gamma\gamma$      & SM\ 
Higgs,  120$< M_{H}<160$~GeV \\
$\GG\to H \to t\overline{t}$      & SM\ 
Higgs,  $ M_{H}>350$ GeV \\
\hline
$\GG \to H,A\to b\bar b$  &
 MSSM\ heavy Higgs,  interm. $\tan\beta$\\
$\GG\to \tilde{f}\bar{\tilde{f}},\
\tilde{\chi}^+_i\tilde{\chi}^-_i$ & large cross sections \\ 
$\GG\to \tilde{g}\tilde{g}$ & measurable cross sections\\
$\GG\to  H^+H^-$ & large cross sections \\ 
$\GG\to S[\tilde{t}\bar{\tilde{t}}]$ & 
$\tilde{t}\bar{\tilde{t}}$ stoponium  \\
$\GE \to \tilde{e}^- \tilde{\chi}_1^0$ & 
 $M_{\tilde{e}^-} < 0.9 \times 2E_0 - M_{\tilde{\chi}_1^0}$  \\
\hline
$\gamma\gamma \to \gamma\gamma$ &   non-commutative theories \\
$e\gamma \to eG $ &   extra dimensions\\
$\gamma\gamma  \to \phi$    &   Radions \\
$e\gamma \to \tilde{e}\tilde{G} $ & superlight gravitions\\
\hline 
$\GG\to W^+W^-$ & anom. $W$ inter., extra dimensions \\
$\GE\to W^-\nu_{e}$ & anom.$W$ couplings \\
$\GG\rightarrow 4W/(Z)$& $WW$ scatt., 
quartic anom.~$W$,$Z$\\
\hline
$\GG\rightarrow t\bar{t}$ & anomalous top quark interactions \\
$\GE\rightarrow \bar t b \nu_e$ & anomalous $W tb$ coupling \\
\hline
$\GG\rightarrow$ hadrons & total $\GG $ cross section \\
$\GE\rightarrow e^- X$, $\nu_{e}X$ & NC and CC structure functions \\ 
$\gamma g\rightarrow q\bar{q},\ c\bar{c}$ & gluon in the photon \\
$\GG\to J/\psi\, J/\psi $ & QCD Pomeron \\
\hline
\end{tabular}
\end{center}
}
\caption{Update of the Gold--plated processes at photon colliders.}

\end{table}


\section*{Acknowledgments}
I would like to thank my co-conveners M. Kr\"amer, M. Krawczyk, 
and S. Maxfield of the ECFA/DESY Future LC study group on 
``$\gamma\gamma $ and $e\gamma$ Collider Physics'', for the two 
years
pleasant collaboration
 and critically reading the manuscript.



\begin{thebibliography}{9}


\bibitem{first}
I.~F. Ginzburg, G.~L. Kotkin, V.~G. Serbo, and V.~I. Telnov,
\newblock {\it Nucl. Instr. and Meth.} {\bf 205}, (1983) 47;\\
I.~F. Ginzburg et~al.,
\newblock {\it Nucl. Instr. and Meth.} {\bf 219},  (1984) 5;\\
V.~Telnov,
\newblock {\it Nucl. Instr. and Meth.} {\bf A294},  (1990) 72;\\
V.~Telnov,
\newblock {\it Nucl. Instr. and Meth.} {\bf A355},  (1995) 3.

\bibitem{pctdr}B. Badelek et al., hep-ex/0108012.

\bibitem{golden} E. Boos et al, {\it Nucl. Instrum. Meth.}
{\bf A472} (2001) 100,
hep-ph/0103090.


\bibitem{phocol} V. Telnov, PHOCOL program, private comm.


\bibitem{circe} T. Ohl, {\it Comput. Phys. Commun.} {\bf 101} (1997) 269,
hep-ph/9607454.

\bibitem{compaz}A.F. \.Zarnecki, 
{\it Acta Phys. Polon.} {\bf B34} (2003) 2741,
hep-ex/0207021 (2002).


\bibitem{moenig} K. M\"onig, J. Sekaric, talks during this workshop.


\bibitem{neutron_telnov} V. Telnov, Talk at the ECFA/DESY meeting
in Amsterdam April 2003.

\bibitem{kraus} F. Kraus, talks during this workshop.

\bibitem{wing} J. Butterworth, M. Wing, talks at ECFA/DESY workshops, 
Prague Nov. 2002 and
Amsterdam, April 2003;\\
J. M. Butterworth, S. Butterworth, hep-ph/0210404, 
{\it Comp. Phys. Comm.} {\bf 153}
(2003) 164.



\bibitem{moenig_lum} V. Makarenko, K. M\"onig, T. Shishkina,  
LC-PHSM-2003-016.

\bibitem{telnov_lum} A.V. Pak et al., LC-PHSM-2003-057.


\bibitem{wwa} V. Budnev { et al.}, {\em Phys.\ Rep.\ } {\bf C15}
 181 (1975). 






\bibitem{pancheri}R. M. Godbole, G. Pancheri
{\it Eur. Phys. J. } {\bf C19}  (2001) 129, hep-ph/0010104 


\bibitem{kwiecinski} N. Timmeanu, J. Kwiecinski and L. Motyka, 
{\it Eur. Phys. J.} {\bf C23} (2002) 513,
hep-ph/0110409.

\bibitem{jankowski} F. Cornet, P. Jankowski, M. Krawczyk and A. Lorca,
{\it Phys. Rev.} {\bf D68} (2003) 014010, hep-ph/0212160.

\bibitem{total}R.M. Godbole, A. De Roeck, A. Grau, G. Pancheri 
{\it JHEP} {\bf0306},  (2003) 61, hep-ph/0305071.

\bibitem{soldner} S. Soldner-Rembold and G. Jikia,
{\it Nucl. Instrum. Meth.} {\bf  A472} (2001) 133, hep-ex/0101056.


\bibitem{piotr}P. Niezurawski, A.F.\.Zarnecki, M. Krawczyk, hep-ph/0307183 
and  LC-PHSM-2003-086;\\
P.Niezurawski, A.F.\.Zarnecki, M.Krawczyk, 
          {\it  Acta Phys. Polon. } {\bf B34} (2003) 177, hep-ph/0208234.


\bibitem{rosca} A. Rosca and K. M\"onig, LC-PHSM-2003-084.

\bibitem{jikia}
G. Jikia and A. Tkabladze, {\it Nucl. Inst. Meth} {\bf A355}, (1995) 91;
{\it Phys. Rev.} {\bf D54}  (1996) 2030, hep-ph/9406428.



\bibitem{zarnecki} P. Niezurawski, A.F.\.Zarnecki, M. Krawczyk, hep-ph/0307175 
and LC-PHSM-2003-088;\\
P.Niezurawski, A.F.\.Zarnecki, M.Krawczyk, {\it JHEP}
 0211 (2002) 034, hep-ph/0207294.


\bibitem{mariax} See e.g.. 
I. Ginzburg, M Krawczyk and P. Osland 
 LC-PHSM-2003-037 and references therein.






\bibitem{godbole}R. M. Godbole, S. D. Rindani, R. K. Singh,
{\it Phys. Rev.} {\bf D67} (2003) 095009, hep-ph/0211136.

\bibitem{asner2} D. Asner et al, 2002, hep-ph/0208219.

\bibitem{asner1} D. Asner et al 2003, hep-ph/0308103.

\bibitem{asner}D. Asner, J. Gronberg, J. Gunion, 
{\it Phys. Rev.} {\bf D67} (2003) 035009,  
hep-ph/0110320.

\bibitem{zarnecki2} P. Niezurawski, A.F.\.Zarnecki, M. Krawczyk,
hep-ph/0307180 and  LC-PHSM-2003-088.


\bibitem{moenig_tgc}K. M\"onig, J. Sekaric, LC-PHSM-2003-072, I. Marfin,
V. Mossolov, T. Shishkina, LC-PHSM-2003-085.

\bibitem{whizard} W. Kilian, Procs of the Int.  Conf. on High Energy
Physics (ICHEP 2002), Amsterdam, p831.



\bibitem{godbole_dip} 
R.M. Godbole, P. Poulose and S.D. Rindani,
LC Note to appear.


\bibitem{watanabe}
I. Watanabe et al., KEK-REPORT-97-17 (1998).

\bibitem{berge}Stefan Berge, Michael Klasen, hep-ph/0303032.

\bibitem{rizzo}T. Rizzo, {\it Nucl. Instrum. Meth.} XS{\bf A472}
 (2001) 37, hep-ph/0008037.



\bibitem{godbole_add}R.M. Godbole, P. Mathews and S.D. Rindani,
LC Note to appear.


\bibitem{berlin} G. Klemz et al., talks during this workshop.

\bibitem{gronberg} J. Gronberg, talks at meetings during this workshop;\\
A.J. Bayramian, R.J. Beach, W. Behrendt, C. Bibeau, et al.,
Tech. digest OSA TOPS on Advanced
      Solid-State Lasers, Vol. 83, pp. 268-275.


\bibitem{jejuheusch} C. Heusch, Proc. of LCWS02, p763, (2002).

\bibitem{schreiber} S. Schreiber, {\it Int. J. Mod. Phys.} 
{\bf A18} (2003) 2827.

\bibitem{gross} E. Gross, Proc. of LCWS02, p83, (2002);\\
M. Krawczyk, hep-ph/0307314.


\end{thebibliography}
\end{document}